# Room temperature one-dimensional polariton condensate in a ZnO microwire

Liaoxin Sun,[1,3] Shulin Sun,[1] Hongxing Dong,[1] Wei Xie,[1] M. Richard,[2] Lei Zhou,[1] Zhanghai Chen,[1,a] L. S. Dang,[2] Xuechu Shen[1,b]

[1] *State Key Laboratory of Surface Physics and Department of Physics, Fudan University, Shanghai 200433, China*

[2] *NPSC, CEA/CNRS/UJF, Institut Néel, 25 Avene des Martyrs, 38042 Grenoble, France*

[3] *National Laboratory for Infrared Physics, Shanghai Institute of Technical Physics, Chinese Academy of Sciences, 500 Yu Tian Road, Shanghai 200083, China*

(Dated: July 27, 2010)

[a] Electronic address: zhanghai@fudan.edu.cn
[b] Electronic address: xcshen@fudan.edu.cn

**A cavity-polariton, formed due to the strong coupling between exciton and cavity mode, is one of the most promising composite bosons for realizing macroscopic spontaneous coherence at high temperature [1-6]. Up to date, most of polariton quantum degeneracy experiments were conducted in the complicated two-dimensional (2D) planar microcavities. The role of dimensionality in coherent quantum degeneracy of a composite bosonic system of exciton polaritons remains mysterious. Here we report the first experimental observation of a one-dimensional (1D) polariton condensate in a ZnO microwire at room temperature. The massive occupation of the polariton ground state above a distinct pump power threshold is clearly demonstrated by using the angular resolved spectroscopy under non-resonant excitation. The power threshold is one order of magnitude lower than that of Mott transition. Furthermore, a well-defined far field emission pattern from the polariton condensate mode is observed, manifesting the coherence build-up in the condensed polariton system.**

A significant progress towards observation of a coherent macroscopic quantum state in semiconductor systems has been made in the last several years [3-6]. Experimental evidence for the Bose-Einstein condensation of exciton-polaritons [4], and subsequently quantized vortices, Bogoliubov dispersion and superfluidity of such coherent polariton condensate were observed in semiconductor microcavities [7,8,9]. Very recently, polariton lasing, i.e. coherent light emission originating from the polariton condensation is also demonstrated in microcavities made of GaN and organic materials [10,11]. Up to date, most of the experimental observations of polariton quantum degeneracy were carried out in two-dimensional (2D) planar microcavities where the light field was confined in the growth direction. Although one step towards one dimensional (1D) cavity polaritons has been taken very recently, and manifold channels for parametric wave mixing and extended condensate in 1D system were reported at low temperature [12,13], the role of dimensionality in coherent quantum degeneracy of composite bosonic system of exciton polariton remains unexplored. Moreover, besides its great importance in fundamental physics, realizing coherent polariton condensates in a 1D structure is highly desirable for "polaritonics" device applications. 1D cavity polaritons are an ideal system for fabricating single-mode polariton lasers and entangled polariton pair sources for quantum communication. From the point of view of application, realizing polariton devices operating at elevated temperatures (room temperature or above) is essential. In the pioneering work on the 1D polariton systems [12,13], the structures were etched from a 2D GaAs planar microcavity which requires elaborated nano fabrication processes, and the weak exciton binding energy of GaAs makes it impossible for a room temperature device application. So far, the macroscopic spontaneous coherence of 1D cavity polaritons and 1D polariton lasing at room temperature are yet to be demonstrated. From both fundamental physics and device application points of view, it is crucial to explore a simple but effective method for fabricating 1D microcavities based on wide bandgap semiconductors (whose exciton binding energies and exciton-photon coupling strength are large enough for the stable existence of excitonic polaritons). ZnO nanowires (or microwires) functioning as whispering gallery (WG) microcavity fulfill the requirements of polariton stability at room temperature and relatively simple fabrication process. Recently, the strong exciton-cavity-mode coupling has been reported at room temperature in these structures [14]. The geometry of microwire provides a 2D confinement for polaritons in the cross section plane. In such a simple system, the overlap between excitons and electromagnetic field can be close to unity, giving rise to a Rabi splitting of hundreds of meV. Furthermore, owing to this extremely large Rabi splitting, the decoupling of polaritons with LO phonons is largely enhanced and thus the dephasing of polaritons is greatly suppressed. This is favorable for realizing room temperature polariton quantum degeneracy [15]. In this Letter, we present the first observation of the spontaneous coherence of 1D polaritons in a simple ZnO whispering gallery microcavity at room temperature.

ZnO microwires with 1.45 μm radius and over 100 μm length are fabricated by using a simple oxidation-sublimation process. A typical scanning electron microscopy (SEM) image of a single ZnO microwire is shown in Fig. 1(a). It can be seen that the synthesized microwire has very smooth and flat surfaces and inerratic hexagonal cross sections. This ensures that the cross sections of the wire form natural WG microcavities with very high quality factors. Since a WG microcavity provides two-dimensional confinement of photons (and polaritons), this uniform wire-like structure is actually a 1D polariton system characterized by free wave vector of cavity polariton along the c-axis of ZnO microwire. To confirm this, we performed the dispersion measurements by using the angular resolved photoluminescence spectroscopy with a continuous wave excitation at 325 nm by a He-Cd laser in two optical configurations. Observing the photoluminescence from the structure angularly resolved with respect to the c-axis of microwire, we obtained the dispersion of polariton energy versus the wave vector $k_z$ (E=E($\theta_z$)). We have determined the polariton dispersion in the confined 2D plane scanning the azimuthal observation angle (E=E($\theta_{x,y}$)). We have studied both TE (the electrical component of light

E⊥c-axis) and TM (E∥c-axis) polarized modes of the WG cavity.

Figure. 1(b), (c) displays the angular resolved emission mapping results for configurations of E=E($\theta_z$) and E=E($\theta_{x,y}$), respectively, obtained using the excitation spot size less than 3 μm$^2$. In Fig. 1(b), modes with parabolic-like dispersions are visible in the spectral region of green-blue light. These modes show very narrow linewidths (~ 3 meV) and the estimated Q factor is as high as 1000, which is the largest value ever been obtained in ZnO microcavities. Close to the energies of the bare excitons (A-exciton: 3.309 eV; B-exicton: 3.315 eV; C-exciton: 3.355 eV), the curvatures of mode dispersions gets smaller and smaller, and the modes clearly show the repulsion-like behavior at large angles. These behaviors demonstrate the coupling between optical cavity modes and excitons. Due to the optical selection rules, the TE and TM polarized WG modes would interact with different exciton states (A and B excitons are mainly coupled to TE modes while C excitons interact with TM modes). Thus, these polariton dispersions can be grouped into two classes characterized by TE and TM polarizations, respectively. Fig. 1 (c) shows the polariton dispersion in the plane normal to the c-axis of the microwire, the bright straight lines in the figure indicate the dispersionless behavior of polariton modes, which can be easily understood considering the 2D confinement in the plane. This behavior is a fingerprint of the 1D cavity polaritons. The Rabi splitting is extracted by fitting the experimental dispersions, based on the simple coupled oscillator model. Unlike the 2D case, the anisotropy of the excitonic -light coupling must be taken into account for WG cavity polaritons in microwires. The results of simulation are shown by color-dotted curves in Fig. 1(b). One can see that the theory reproduces reasonably well with the measured polariton dispersions. A more detailed and complex analysis of the dispersions

by solving analytically Maxwell's equations involving polariton and anisotropic effect can be found in Ref (15). From the best theoretical fitting to the experimental results, we obtained the Rabi splitting for 1D cavity polaritons in ZnO larger than 300 meV. This Rabi splitting and the corresponding figure of merit as high as 100 are the largest values ever observed in semiconductor microcavities to the best of our knowledge.

The high Q factor and the remarkably large Rabi splitting make 1D ZnO microcavities excellent candidates for the experimental observation of the nonlinear behavior of 1D cavity polaritons at room temperature. To study the non-linear optical properties of these structures, we used an Nd: YAG pulse laser with the output wavelength of 355 nm and the pulse repetition rate of 10 Hz as excitation capable of creating high carrier densities in the microcavity. Here, we focus on the spectral range of Fig. 1, where the polariton modes $N_{TE}$=46, 47, 48, 49 can be observed. In Fig. 2 (a), the angular resolved emission spectra obtained at different pump powers are displayed. It can be seen that with the average pump power increasing, stimulated emission first occurs to the polariton mode $N_{TE}$=48. The sharp crossover characterized by a nonlinear increase of the emission intensity for the polaritons with $k_z$=0 (i.e. the ground-state polaritons) appeared at 30 nW. At the threshold power $P_{th}$, the estimated polariton density $N_{3D}$ is about $4 \times 10^{18} cm^3$, which is one order of magnitude below the Mott density $\approx 5.5 \times 10^{19} cm^3$ for ZnO. This implicitly confirms the existence of the strong coupling regime above the power threshold. Furthermore, for clarity, the integrated intensities of polariton ground-state emission are shown in Fig. 2 (b) and (c) as functions of pump power and the population distribution along the polariton dispersion above the threshold, respectively. The stimulated emission behavior is well demonstrated by the fact that the integrated intensity is enhanced by two orders of magnitude as the average power increases from 30 nW to 100 nW, whereas it increases slowly below the threshold. Above threshold, the power-dependent intensity can be well described by the power law $P^{3.90 \pm 0.27}$, which indicates a superlinear increase characteristic for lasing. In contrast to the behavior below threshold, the polariton distribution in $k_z$-space above the threshold shows the massive occupation at the bottom of the parabolic-like polariton dispersion, i.e. the polariton ground state at $k_z=0$. These results imply that, regardless of the system being at equilibrium or out of thermal equilibrium, the polaritons are likely to be condensed at the lowest energy state in 1D $k$-space above threshold. Here, we note that the origin of this nonlinear phenomenon is mainly due to the stimulated relaxation of polariton-polariton or polariton-phonon scattering, which should be distinguished from the conventional laser based on the emission amplification due to the population inversion. This spontaneous emission of light by a macroscopically occupied polariton state is usually called "polariton laser" [2]. We know that the relaxation process strongly depends on the detuning between excitons and the pure optical modes (determined by the component ratio of exciton and photon in a polariton). As reported in Ref. (15), the minimum pump power threshold preferentially appears at the positive detuning regime in a CdTe microcavity, which is determined by interplay between thermodynamic and kinetic factors. In our experiment, the similar tendency is observed. By comparing with the fitting results, we obtained a positive detuning of 50 meV for the polariton condensate mode $N_{TE}$=48 which has the lowest power threshold. As the pump power increases to 80 nW, the polariton mode $N_{TE}$=47, with the detuning of -30 meV at $k_z$=0, begins to show a similar nonlinear enhancement of the emission intensity near the ground state.

With increasing the pump power, three polariton dispersions with mode numbers of $N_{TE}$=46, 47, 48 show the obvious blueshifts in energy, and the broadening of the emission lines at $k_z$=0 (θ=0) above the threshold is observed. Figure. 3 displays the blueshifts of the three polariton modes at $k_z$=0 as functions of the pump power. For the polariton mode with $N_{TE}$=48, it can be seen that the ground-state energy is blueshifted by 11 meV when the pump power is increased from 10.5 nW to 69.5 nW. This blueshift is fairly small compared to the very large Rabi splitting of ~300 meV in our sample. This indicates that the polariton condensate occurs in the strong coupling regime. When the pump power passes through the threshold of 30 nW, the $N_{TE}$=48 polariton dispersion shows an abrupt blueshift, as one can see in Fig. 3(a). However, what is more interesting is that for the modes $N_{TE}$=46 and 47, the similar behavior is also observed even though they remain in the non-condensate regime. This result indicates that the blueshift of polariton modes seems to be weakly dependent on the occupation of particular modes but reflects the total polariton concentration in the system. The blueshift of the dispersion is usually ascribed to the polariton-polariton repulsive interactions, and may be also affected by the decrease of the exciton oscillator strength due to the phase space filling. The interactions are likely to play a dominant role in the blueshift far from the Mott transition limit [17]. Although each polariton mode is an individual eigenstate of the ZnO microwire system, they are all formed by the same excitons coupled with the different cavity modes. Thus each polariton mode exhibits similar blueshift behavior as the pump power increased.

Above the threshold, two polariton dispersions (guided by dotted (*L* mode) and dashed (*NL* mode) curves in Fig. 2(a)) were observed simultaneously in the spectral mapping. To explain this phenomenon, we explored the dynamical recombination emission process of the polaritons. In our experiment, the repetition rate and the pulse duration of the excitation is 10 Hz and 3 ns, respectively. The time spacing between two adjacent pulses is 100 ms, which is $10^8$ longer than the lifetime of polaritons. Therefore, one can imagine that the massive occupation at the polariton ground-state occurring above the threshold would go quickly back in the non-condensate regime because of the decrease of the polariton density without continuous supplement. Considering this dynamical process and the long integrated time for the emission collection, the observed "*L*" mode and "*NL*" mode can be identified as the polariton dispersion in the presence of the condensate and non-condensate dispersions of polaritons, respectively.

So far, we have been focusing on the evolution of the polariton energy dispersions versus $k_z$ along the c-axis (E=E($\theta_z$)) with the pump power increasing. The massive occupation at the polariton ground-state in $k_z$-space has resulted in the bosonic final state stimulation of the polaritons scattering (Fig. 2(c)). In the following, we explore the evolution of polariton dispersion versus $k_{x,y}$ (E=E($\theta_{x,y}$)) above the threshold. Fig. 4 (a) displays the polariton dispersion in the plane normal to the c-axis of ZnO microwire above the condensation threshold. Surprisingly, we observed that the emission intensity along the in-plane polariton dispersion curve exhibited the far field standing wave-like emission pattern containing six bright spots within about ±30° angle range. This phenomenon is not observed below the threshold, as shown at Fig. 1 (c). This suggests that the appearance of the emission pattern is related to the polariton condensate. We ascribed this phenomenon to the coherence buildup in the polariton condensate system. As we all know, below the threshold, WG exciton polaritons are independent from each other because the phases of polaritons are stochastic. Thus, the overlap of these polariton matter waves could not generate the interference emission patterns at the plane perpendicular to the c-axis of ZnO microwire. However, as the pump power reaches the threshold, the stimulation of polaritons scattered to the ground-state leads to a macroscopic coherent state. In this case, the interference of coherent polaritons exhibits a well defined in-planed intensity pattern, resulting in the far-field directional emission patterns appearing in the angular resolved photoluminescence spectra. To understand this phenomenon, we performed simulations based upon the finite-element method (FEM) to study the laser eigen-modes wave properties in the ZnO hexagonal WG microcavity. In our simulations, the exciton-photon coupling is considered as a contribution to the refractive index of the medium [18] and the gain effect of the active medium is carefully taken into account. Near the wavelength of ~388 nm, we carefully identified from simulations the possible $N_{TE}$=48 polariton condensate mode observed experimentally and depicted the field distribution in this mode in Fig. 4 (b). The pattern shows clearly that the emission from the WG microcavity is highly directional with six bright emission spots appearing in the about ±30° angular range (labeled as solid lines and dashed circles) in the excellent agreement with experimental observations. In addition, we performed the double-slit interference experiment, in which the separation of two slits is chosen to be equal to the distance between bright spot 2 and 5 (Fig. 4(b)). We can see a pronounced modulation of the intensity distribution above the power threshold due to interference, which confirms the coherence of our polariton system. We conclude that the coherence build-up in the polariton condensate system is responsible for the appearance of in-plane emission pattern.

In summary, we have reported for the first time the 1D polariton condensate in ZnO WG

microcavity at room temperature. The excitation threshold for polariton lasing is determined from the measured nonlinear enhancement of the emission intensity. A significant shrinking of the polariton emission to the $k=0$ at the threshold manifesting the i.e. massive occupation of the polariton ground-state is clearly observed. The directional polariton laser effect is accompanied by the spatial coherence build-up in the polariton system, above threshold which is well verified by the double-slit interference experiment.

**Methods**

**Sample preparation,** The growth of the ZnO microwires was carried out in a horizontal tube furnace. No catalysts, carrier gas, low pressure, or templates were used in the experimental procedure. A mixture of ZnO powders and graphite powders with weight ratio of 1:1 was loaded into a small quartz boat. A clean Si wafer was covered over the top of the boat, and then the boat was placed in the center of the quartz tube. The temperature of the tube furnace was raised to 1000℃ at a rate of 25℃/min and the temperature was maintained at 1000℃ for 60 min. After the furnace is cooled down to room temperature, a large amount of ZnO microwires was found in the Si wafer substrate. For the micro-photoluminescence measurements of a single microwire, we first disperse the microwires in ethanol by sonicating and then transfer them onto a transparent quartz slide. A single microwire with uniform diameter and high mode quality factor, which is necessary for angular resolved (momentum resolved) spectroscopy investigation, can be chosen via careful spectra measurements.

**Experiment setup,** spatial- and angular-resolved spectroscopic system is build up by using UV enhanced objective (40X) and PI monochrometer equipped with Si CCD. The fourier plane image is carefully projected onto the entrance slit of monomchromator. A He-Cd continuous wave laser with output wavelength of 325 nm is used as an excitation source, which is focused onto the microwire with excitation spot of about 3 μm$^2$ by 40X UV objective. For spectral investigations of polariton spontaneous coherence in ZnO microwire, the other kind of excitation source, Nd: YAG pulse laser with output wavelength of 355 nm, is used. The laser duration time and repetition rate is 3 ns and 10 Hz, respectively.

Acknowledgments

The work is funded by the NSF and 973 projects of China (No. 2004CB619004 and No. 2006CB921506).

[1] C. Weisbuch, M. Nishioka, A. Ishikawa, &Y. Arakawa, Observation of the coupled exciton-photon mode splitting in a semiconductor quantum microcavity, *Phys. Rev. Lett.* **69**, 3314 (1992).

[2] A. Imamoglu, R. J. Ram, S. Pau and Y. Yamamoto, Nonequilibrim condensates and lasers without inversion: Exciton-polariton lasers, *Phys. Rev. A*. **53**, 4250 (1996).

[3] H. Deng, *et al.*, Condensation of semiconductor microcavity exciton polartion. *Science* **298**, 199 (2002).

[4] J. Kasprzak, *et al.*, Bose-Einstein condensation of exciton polaritons. *Nature* **443**, 409 (2006).

[5] R. Balili, V. Hartwell, D. Snoke, L. Pfeiffer & K. West, Bose-Einstein condensation of microcavity polaritons in a trap, *Science* **316**, 1007 (2007).

[6] Daniele Bajoni, *et al.*, Polariton laser using single micropillar GaAs-GaAlAs semiconductor cavities, *Phys. Rev. Lett.* **100**, 047401 (2008).

[7] K. G. Lagoudakis, *et al.*, Quantized vortices in an exciton-polariton condensate, *Nature Phys.* **10**, 1038 (2008).

[8] S. Utsunomiya, *et al.*, Observation of bogoliubov excitations in exciton-polariton condensates,

*Nature phys.* **10**, 1034 (2008).

[9] A. Amo, *et al.*, Collective fluid dynamics of a polariton condensate in a semiconductor microcavity, *Nature*, **475**, 291-295 (2009).

[10] S. Christopoulos, *et al.*, Room temperature polariton lasing in semiconductor microcavities, *Phys. Rev. Lett.* 98, 126405 (2007).

[11] S. Kéna-Cohen, and S. R. Forrest, Room-temperature polariton lasing in an organic single-crystal microcavity, *Nature Photon.* **10**, 1038 (2010).

[12] G. Dasbach, M. Schwab, M. Bayer, D. N. Krizhanovskii, & A. Forchel, Tailoring the polariton dispersion by optical confinement: Access to a manifold of elastic polariton pair scattering channels, *Phys. Rev. B.* **66**, 201201(R) (2002).

[13] E. Wertz, et al., Spontaneous formation and optical manipulation of extended polariton condensates, arXiv: 1004. 4084v1

[14] Liaoxin Sun, et al., Direct observation of whispering gallery mode polaritons and their dispersion in a ZnO tapered microcavity, *Phys. Rev. Lett.* **100**, 156403 (2008).

[15] A. Trichet, L. Sun, G. Pavlovic, N. A. Gippius, G. Malpuech, W. Xie, Z. Chen, M. Richard and Le Si Dang, arXiv: 0908.3838v2.

[16] J. Kasprzak, D. D. Solnyshkov, R. André, Le Si Dang, & G. Malpuech, Formation of an exciton polariton condensate: thermodynamic versus kinetic regimes, *Phys. Rev. Lett.* **101**, 146404 (2008).

[17] E. del Valle, *et al.*, Dynamics of the formation and decay of coherence in a polariton condensate, *Phys. Rev. Lett.* **103**, 096404 (2009).

[18] Comsol Multiphysics by COMSOL ©, ver. 3.5, network license(2008)

**Figure Captions:**

Figure 1: (a) The scanning electron microscopy of single ZnO microwire. The angular resolved (k-space) emission mapping of exciton polaritons dispersion both in the plane parallel to the c-axis of ZnO microwire (b) and in the plane perpendicular to the c-axis of ZnO microwire (c).

Figure 2: (a) The evolvement of exciton polariton dispersions ($E=E(\theta_z)$) as a function of pump power (logarithmic plots of the intensity). The massive occupation of ground-state for polariton mode $N_{TE}$=48 appeared when the pump power higher than the threshold $P_{th}$=30 nW. (b) The emission intensity of polariton ground-state shows a superlinear increasing, which can be fitted by the power law $P^{3.90 \pm 0.27}$, above the threshold. (c) The polariton occupancy distribution along the polariton dispersion above the threshold with pump power 69.5 nW (olive sphere) under Nd: YAG pulse laser excitation and below the threshold (red sphere) under He-Cd laser excitation.

Figure 3: The blueshift of polariton ground-state energy with the pump power increasing for polariton modes $N_{TE}$=48, 47, 46. The threshold of polariton condensate for $N_{TE}$=48 is label with dashed line.

Figure 4: (a) The standing-wave-like emission pattern appeared above the threshold in the in-plane polariton dispersion measurement for the polariton mode $N_{TE}$=48. (b) The double-slit interference pattern and the simulated emission intensity pattern of polariton condensate mode $N_{TE}$=48 by using the FDTD method.

Figure 1

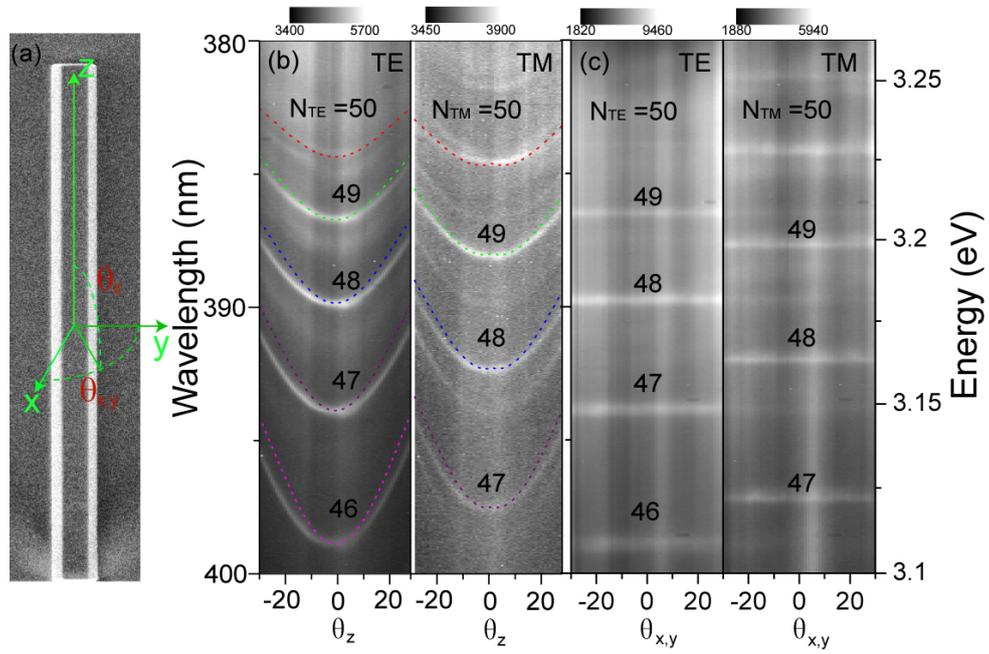

Figure 2

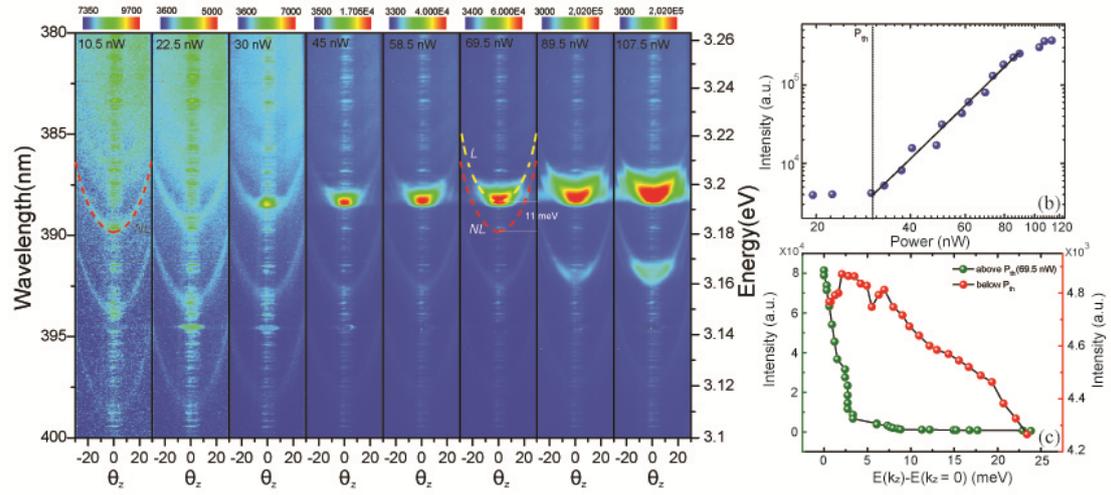

Figure 3

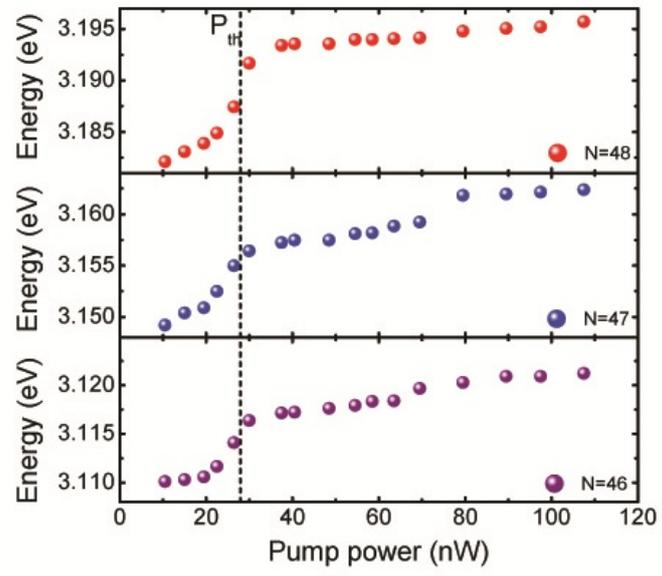

Figure 4

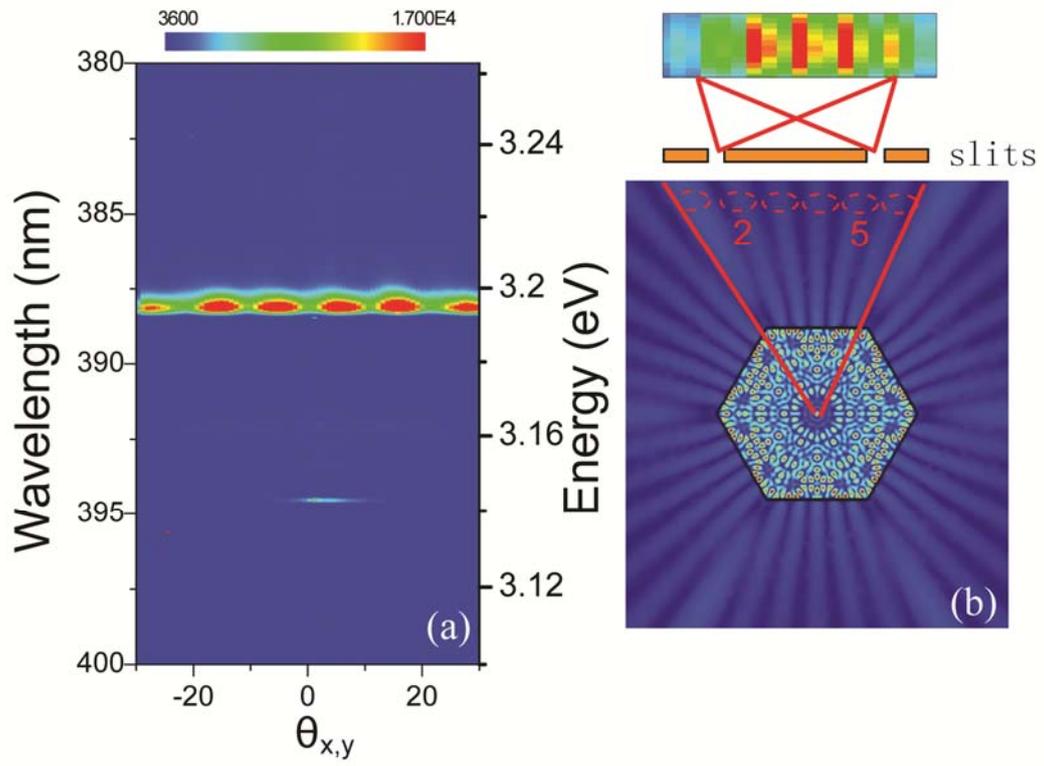